\documentstyle [12pt]{article}
\begin{document}
\newcommand{\vm}{\vspace{0.2cm}}
\newcommand{\vl}{\vspace{0.4cm}}

\title{Why some stars seem to be older than the Universe? }
\author{Matti Pitk\"anen\\
Torkkelinkatu 21 B 39, 00530, Helsinki, FINLAND}
\date{6.2. 1995}
\maketitle
\newpage

\begin{center}
Abstract
\end{center}

\vm

There is some experimental evidence that some stars are older than the
 Universe in General Relativity based cosmology.  In TGD based cosmology
 the paradox has explanation.  Photons can be either topologically
condensed on background spacetime surface or  in 'vapour phase'  that is
progate in $M^4_+\times CP_2$ as small surfaces. The time for  propagation
from A to B is  in general larger in condensate than in vapour phase.  In
principle observer detects from a given astrophysical object both vapour
phase and condensate photons,  vapour phase photons being emitted  later
than condensate photons.   Therefore the   erraneous identification of
 vapour phase photons as condensate photons leads to an  over estimate for
the age of the star and star  can look  older than the Universe.  The
Hubble constant for vapour phase photons is that associated with $M^4_+$
and smaller than the  Hubble constant of matter dominated cosmology. This
could  explain the measured two widely different values of Hubble constant
if  smaller Hubble constant  corresponds to the Hubble constant of the
future light cone $M^4_+$. The  ratio of propagation velocities of  vapour
phase and condensate photons  equals to the  ratio of the two Hubble
constants, which in turn is depends on the ratio of mass density and
critical mass density, only.  Anomalously large redshifts are possible
since vapour phase photons can come also from region outside the horizon.

\newpage

\section{Why some stars  seem to be older than the  Universe?}

There exists experimental evidence that some stars are older than
 Universe \cite{Pierce,Freedman,Saha}. A related problem is
the problem of two Hubble constants. These
paradoxical results can be understood in TGD:eish cosmology.   In TGD light
 can propagate in two manners.  In  topological
 condensate  light ray propagates along curved spacetime surface as a
small condensed particle and in vapour phase as a small 3-surface in
imbedding space $H=M^4_+\times CP_2$,  where $M^4_+$ is future light cone
of $M^4$.  The time needed to travel from point A to point B is shorter in
vapour phase since the geodesic length along spacetime surface  in
the  induced metric is obviously longer than in free Minkowski
space. The failure to regard vapour phase photons as condensate photons
 leads to the paradox as following arguments shows and also to a problem of
two different Hubble constants.
Moreover, the possibility  of vapour phase photons emitted by objects
 outside the spacetime horizon explains  also objects with anomalously
large redshifts.

\vm

To understand these results  one must study
TGD:eish cosmology in more quantitative level.
 \\ a) The most general cosmological imbedding of
$M^4_+$ to
 $M^4_+\times CP_2 $,   is of form

\begin{eqnarray}
s^k &=& s^k(a)\nonumber\\
g_{aa} &=& 1-s_{kl}\frac{ds^k}{da}\frac{ds^l}{da}\nonumber\\
ds^2&=& g_{aa}da^2- a^2(\frac{dr^2}{1+r^2}+r^2d\Omega^2)
\end{eqnarray}

\noindent Here $s_{kl}$ is $CP_2$ metric tensor and describes always
expanding cosmology with subcritical or at most critical mass density
\cite{TGD}. \\
b)   The age of the Universe detefined as $M^4_+$  proper time $a$ of the
comoving observer (comoving observer on spacetime surfaces is also
 comoving in
$M^4_+$)   is  larger than  the age defined as the  proper time  $ s (a)$
of
the comoving
 observer on spacetime surface.  For the  matter dominated Universe   one
has
$g_{aa}=Ka $  so that
one has

\begin{eqnarray}
\frac{age(cond)}{age(vapour)}&=& \frac{s(a)}{a}= \frac{2}{3} \sqrt{g_{aa}}
\end{eqnarray}

\noindent for the ratio of the ages.\\
c) $g_{aa}$ can estimated from the
expression for mass density in expanding cosmology

\begin{eqnarray}
\rho&=& \frac{3}{8\pi G}(\frac{1}{g_{aa}}+k)\nonumber\\
k&=&-1
\end{eqnarray}

\noindent $k=0$ mass density corresponds to critical mass density $\rho_c$.
 The mass
density is believed to be a fraction of order $\epsilon= 0.1-0.5$ of the
critical mass density and this gives  estimate for  $\sqrt{g_{aa}}$:

\begin{eqnarray}
\sqrt{g_{aa}}&=&\sqrt{1-\epsilon}\nonumber\\
\epsilon&=& \frac{\rho}{\rho_c}
\end{eqnarray}

\noindent $\sqrt{g_{aa}}=2/3$ suggested by the proposed solution to Hubble
constant discrepancy gives $\epsilon=\frac{9}{4}$. $\epsilon=.1$ gives
$\sqrt{g_{aa}}\simeq .95$.
\\ d) The
 ratio of the condensate travel time to vapour phase travel time
for short distances  is given by

\begin{eqnarray}
\frac{\tau (cond) }{\tau (vapour)} &=& \frac{1}{\sqrt{g_{aa}}}
\end{eqnarray}

\noindent This effect is in principle observable and the considerations of
\cite{TGD} suggest that $g_{aa}$ can differ from unity by a factor as
 large as
one half. The effect provides also a means of measuring the mass density
of the Universe.
 \\ e) The light travelling in vapour phase can reach observer  from a
region, which  is the intersection of the past light cone of the observer
 with
the  boundary of  $M^4_+$ and therefore finite region of $M^4$.   The $M^4$
radius of this region in the rest frame of the observer is
equal  $r_M= a/2$ by elementary geometry. \\
f) For a null geodesic  of spacetime surface starting at
 $(a_0,r)$ and ending at $(a,0)$
 one has

\begin{eqnarray}
r&=& sinh(\int_{a_0}^a\frac {\sqrt{g_{aa}}}{a} da)
\end{eqnarray}

\noindent If $g_{aa}$ approaches zero for $a_0\rightarrow 0$ as it does
 for
 radiation dominated cosmology the integral on the right hand side is
finite.
This
means that the value of $r_M(a_0)$ ($M^4$ distance of the object from
 observer) approaches zero at this limit. All radiation
from the moment of big bang comes from the dip of the light cone.
 In TGD the Planck time cosmology with critical mass density corresponds to
$g_{aa}= K$, $K$  very small number and also in this case the radiation
comes
from origin.\\ g)  The radius $r_M(a_0)$ has maximum for some finite value of
$a_0$ and this
 radius defines the $M^4$  radius of the  Universe observed by condensate
 photons. The maximum
corresponds to rather large value of $a_0$ so that  one can approximate the
cosmology with matter dominated cosmology:  $g_{aa}= Ka$ and one has
the condition

\begin{eqnarray}
u_0&=&2tanh(u-u_0)\nonumber\\
u&=& \sqrt{Ka}\\
u_0&=& \sqrt{Ka_0}
\end{eqnarray}

\noindent  The following table gives the values of $\sqrt{\frac{a_0}{a}}$
and
$\frac{r_M(max,cond)}{r_M(max,vapour)}$ for $\sqrt{g_{aa}}=\sqrt{Ka}=2/3$
 and
$1$ respectively.

\vl

\begin{tabular}{||c|c|c||}\hline\hline
$\sqrt{Ka}$&$\sqrt{\frac{a_0}{a}}$&$\frac{r_M(max,cond)}{r_M(max,vapour)}$
\\ \cline{1-3}
\hline
 $\frac{2}{3}$&.663&.2 \\ \hline
1& .658&.3 \\ \hline\hline
\end{tabular}

\vm

\noindent Note that  anomalously large
redshifts are possible for vapour phase  photons emitted by comoving
objects
outside the horizon and $r_M\ge r_M(max)$. \\
 h) Vapour phase and condensate photons  provide in
principle a possibility to obtain simultaneous information about the
astrophysical object in two different  phases of its development. For
object
situated at distance $r$ and observed at $(a,r=0)$ the emission moments
 $a_0$
and $a_1>a_0$ (in Minkowski proper time) for condensate photon and vapour
 phase
photon are related  by the formula

\begin{eqnarray}
\frac{a}{a_1}&=& exp(2\sqrt{K_1}(a^{1/2}-a_0^{1/2}))
\end{eqnarray}

\noindent in matter dominated cosmology $g_{aa}=K_1a$ ( $K_1a \sim 1$).
Sufficiently nearby Super Nova would  provide a test for this effect. The
 first
burst of light  corresponds to vapour phase photons and second burst to
condensate photons.  The time lag between bursts provides a manner to
 measure
 the
value of $\sqrt{g_{aa}}$.  Unfortunately, the time lag in case of SN1987A
is quite too large since the  distance of order $1.5 \cdot 10^5
\ ly$. The observation of same spectral line with two  different
cosmological redshifts  is second effect of this
kind and might be erraneously interpreted as existence of two different
objects on same line of sight.

\vm

Consider now the solution of the two puzzles.  The previous formula
 explains
 why certain stars seem to be older than the Universe. If one erraneously
identifies vapour phase photons as condensate photons the age of the star
 at
time $a_0<a_1$  is erraneously identified as the age at later time
$a_1$ and this implies that the apparent  age is  given by

\begin{eqnarray}
s(a)_{app}&=&  Xs(a)\nonumber\\
X&=& (\frac{a_1}{a_0})^{3/2}\nonumber\\
\frac{a_1}{a_0}&=& \frac{a}{a_0}exp(-2\sqrt{K}(a^{1/2}-a_0^{1/2}))
\end{eqnarray}

\noindent and larger than the actual age in matter dominated cosmology.
 The apparent ages of lowest stars are roughly by a factor $3/2$ larger
 than the
age of the Universe.  For $\sqrt{Ka}= 1$  and $X=1.57$ this requires
 $s(a_0) \simeq .15s(a)$.  For $\sqrt{Ka}=2/3$ and $X=1.57$ one has
 $s(a_0)\simeq .36
s(a)$.

\vm

 Vapour phase photons provide  a  possible solution to the puzzle of two
 different
Hubble constants if the mass density is sufficiently large. The distances
 derived
from type Ia supernovae give $H_0^a= 54\pm 8 \  km s^{-1} Mpc^{-1}$ to be
compared with the Hubble result $H_0^b= 80 \pm 17 \  km s^{-1} Mpc^{-1}$
\cite{Freedman}. The discrepancy is resolved if the  measurement of  distance
 is
correct and made using vapour phase photons and $H_0^a$ corresponds to the
Hubble constant of $M^4_+$,  which is  by a factor

\begin{eqnarray}
\frac{H_0^a}{H_0^b}&=&
\frac{H_0(M^4_+)}{H_0(X^4)}=\sqrt{g_{aa}}=\sqrt{1-\epsilon} \sim 2/3
\end{eqnarray}

\noindent  smaller than the Hubble constant of spacetime surface.
The needed  mass density $\epsilon=5/9$ and the  ratio
of propagation velocities of  light differs considerably from unity.  For
$\epsilon=.1$ the ratio of two Hubble constants is predicted to be $.95$
and
 some other explanation for discrepancy is needed.

\vl


\begin{thebibliography}{99}
\bibitem[Freedman {\it et al\/}]{Freedman}
Freedman , W.L. {\it et al\/}(1994),Nature 371, 757-762
\bibitem[Pierce {\it et al\/}]{Pierce}
Pierce,M., J {\t et al \/}(1994), Nature 371, 385-389
\bibitem[Saha {\it et al\/}]{Saha}
Saha,A. {\it et al\/} Astrophys. J. (in the press)
\bibitem[Pitk\"anen$_a$]{TGD}
M. Pitk\"anen (1990) {\em Topological Geometrodynamics} Internal
Report HU-TFT-IR-90-4 (Helsinki University).  Summary of
  of Topological Geometrodynamics in book form. Book contains construction
of Quantum TGD, 'classical' TGD and  applications to various branches of
physics. Also TGD inspired  cosmology is discussed.
 \bibitem[Pitk\"anen$_b$]{padTGD}
M. Pitk\"anen (1990) {\em Topological Geometrodynamics and p-Adic Numbers}.
 The book
describes the  general views concerning application p-adic numbers to TGD.
In particular, the manner how  K\"ahler function
defines quantum catastrophe theory is explained.
 \bibitem[Pitk\"anen$_c$]{padmasses}
 M. Pitk\"anen (1994),  {\em
p-Adic description of Higgs mechanism;I,II,III,IV,V}.  hep-th@xxx.lanl.gov
9410058-62. Elementary particle and hadron mass calculations are performed in
parts III,IV.
 \bibitem[Pitk\"anen$_d$]{lightfield}
 M. Pitk\"anen (1994),  {\em p-Adic field theory limit of TGD},
hep-th@xxx.lanl.gov 9412103. Gives general formulation of field theory
 limit in light sector of p-adic TGD. The concept p-adic planewave is
introduced and the absence of ultraviolet divergences is shown.

 \end{thebibliography}
\end{document}